\documentclass[epj,nopacs]{svjour}
\usepackage{latexsym}
\usepackage{graphics}

\begin{document}
\def \ee {\varepsilon}
\thispagestyle{empty}
\title{
On the validity of constraints on light elementary particles
and extra-dimensional physics from the Casimir effect
}

\author{ E.~Fischbach\inst{1},
 G.L.~Klimchitskaya\inst{2,3},
D.E.~Krause\inst{4,1},
V.M.~Mostepanenko\inst{5,3,}\thanks{E-mail:
Vladimir.Mostepanenko@itp.uni-leipzig.de}
}

\institute{
Department of Physics, Purdue University, West Lafayette, Indiana
47907, USA
\and
North-West Technical University, Millionnaya St. 5, St.Petersburg,
191065, Russia
\and
Institute for Theoretical
Physics, Leipzig University, Postfach 100920,
D-04009, Leipzig, Germany
\and
Physics Department, Wabash College, Crawfordsville, Indiana 47933,
USA
\and
Noncommercial Partnership
``Scientific Instruments'',  Tverskaya St. 11, Moscow,  103905, Russia
}
\date{Received: March 5, 2010 / Revised version: date}

\abstract{We discuss the constraints on the parameters of a Yukawa
interaction obtained from the indirect measurements of the Casimir
pressure between parallel plates using the sphere-plate
configuration.
Recently, it was claimed in the literature
that the application of the proximity force approximation 
(PFA) to the 
calculation of a Yukawa interaction in the sphere-plate configuration
 could lead to a large error of order 
100\% in the constraints  obtained. 
Here we re-calculate the constraints both  exactly
and using the PFA, and arrive at identical results. We elucidate
the reasons why an incorrect conclusion was obtained 
suggesting that the PFA is
inapplicable to calculate the Yukawa force.
}

\authorrunning{ E.~Fischbach et al.}
\titlerunning{On the validity of constraints}

\maketitle

The Yukawa-type corrections to Newtonian gravitational law are
predicted from the exchange of light elementary particles and in
some extra-dimensional generalizations of the standard model.
Within a short interaction range of around $0.1\,\mu$m the
strongest constraints on the parameters of Yukawa interaction were
obtained from the indirect dynamic measurement of the Casimir
pressure between two parallel plates using the configuration
of an Au-coated sphere above an Au-coated plate of a
micromachined oscillator \cite{12,11}.
In \cite{12,11} this was done using what is referred to as
{\it the proximity force approximation} (PFA) \cite{2a} for the
calculation of the Yukawa force acting between a sphere and a plate.
For two point-like particles with masses $m_1$ and $m_2$ spaced at a
separation $r$ apart, the Yukawa corrections to Newtonian gravity
are conventionally presented in the
form \cite{3,4}
\begin{equation}
V_{\rm Yu}(r)=-\frac{\alpha Gm_1m_2}{r}\,e^{-r/\lambda},
\label{eq1}
\end{equation}
\noindent
where $G$ is the gravitational constant, $\alpha$ and $\lambda$ are
the strength and the interaction range of the Yukawa interaction,
respectively.

Reference \cite{2} especially investigated the possibility of
calculating the gravitational and Yukawa-type interactions in
sphere-plate configuration using the PFA. In so doing two
formulations of the PFA were presented. In the most general
formulation \cite{2b} the force of interaction is found as
an integral of the known force between parallel surface
elements. According to the simplified formulation of the PFA,
valid under certain conditions, the force is equal to
$2\pi RE$ where $R$ is the sphere radius and $E$ is the energy
per unit area in the configuration of two parallel plates \cite{2a}.
In \cite{2} the applicability of different formulations of the 
PFA for the calculation of Yukawa and gravitational interactions
in the experimental configuration of \cite{12,11} was confirmed. 
Specifically, it was shown that for the gravitational and
Yukawa-type interactions between a compact body and a plane plate
of infinitely large area
the most general formulation of the
PFA leads to the exact results. The simplified formulation
of the PFA was shown to be approximately applicable to the
Yukawa interaction.

Motivated by \cite{2}, the subsequent paper \cite{1} 
discussed the same subject.
In agreement with \cite{2}, paper \cite{1}
arrives at the conclusion
that the level of precision in using the two formulations of the PFA
to evaluate Yukawa forces in the sphere-plane geometry ``is of the same
order of magnitude as the Casimir theory-experiment comparison that uses
PFA to compute the sphere-plate Casimir force.'' On the
basis of \cite{5,6,7,8,9,10} it was also concluded \cite{2}
that for the experimental configuration of \cite{12,11}
with $a/R\approx 0.001$ ($a$ is the closest
separation between a sphere and a plate, $R$ is the sphere radius) the
relative error in the Casimir force due to the use of the PFA is 
about 0.1\%.  The same
conclusion was reached in \cite{1} on the basis of 
\cite{7,8,10a,10b}.
Specifically, \cite{1} states that ``the PFA...{\ }is expected to
approximate the exact Casimir force within 0.1\%.''
Thus, in both papers \cite{2,1} it is recognized that
in recent experiment \cite{12,11} the error in the
evaluation of the Yukawa-type force between a sphere and a plate using
the simplified formulation of the
PFA is of about 0.1\%.

However, notwithstanding the previous discussion, which supports
the use of the PFA, \cite{1} claims that the const\-raints on
$\alpha,\,\lambda$ from the precision dynamic measurements of the
Casimir force obtained in \cite{12,11} could have ``a large
order of 100\% correction.''
According to \cite{1}, ``Considering the relatively small
margins of improvement reported recently (see for instance Fig.~3
in [21]), a systematic shift due to the use of the PFA instead of
EPFA may lead to significant changes for the exclusion region in
the $\alpha-\lambda$ plane.'' We recall that [21] cited
in \cite{1} is just our reference \cite{11} and EPFA is equivalent
to the exact calculation.
By way of contrast, \cite{2} confirmed the
validity of the constraints obtained in \cite{12,11}. Since
these constraints are now
included in the Review of Particle Physics (Particle Data Group)
for the year 2008 \cite{13},
it seems necessary to verify whether the claim of \cite{1}
which casts doubts on
these constraints is correct. In what follows we repeat the derivation of
the constraints
on $(\alpha,\,\lambda)$ from the experimental data of \cite{12,11}
using the exact formulas for the Yukawa-type interaction, and arrive at the
same results as obtained in \cite{12,11} using the 
simplified formulation of the PFA.
We show that the use of the simplified formulation of the  
PFA within its applicability region
for the calculation of Yukawa-type forces cannot lead to any
systematic shifts in the $\alpha-\lambda$ plane not only in
already performed experiments but in
presumably much more precise future
experiments as well. Because of this it is not correct to call
this formulation of
the PFA ``an invalid approximation to compute volumetric
forces'' \cite{1}. As for any approximation, the 
simplified PFA is applicable
in some ranges of parameters (see below) and is not applicable
outside of these ranges. Several other misleading statements of
\cite{1} are also commented upon below.

To begin our derivation, it is pertinent to recall that in the experiment
of \cite{12,11} the separation distance between a sphere of
$R=151.3\,\mu$m radius and a plate
was varied harmonically with time, and the
directly measured quantity was the frequency shift of the oscillator
under the influence of the Casimir force acting between the two bodies.
Due to the properties of the oscillator, this frequency shift is
proportional to the gradient of the Casimir force. 
Using the simplified PFA,
this gradient turns out to be proportional to the effective Casimir pressure
in the configuration of two parallel plates. Thus, in this experiment one
indirectly measures the Casimir pressure between two parallel plates
(see the recent review \cite{15a} for a detailed justification of
this statement).
To obtain constraints on the parameters of Yukawa-type interaction
$\alpha$ and $\lambda$ in \cite{12,11,15b} it was supposed that the
gradient of theYukawa-type force is proportional to the
 Yukawa-type pressure between the two parallel plates, i.e., 
the simplified PFA was
applied to the Yukawa-type forces. As a result, the constraints were obtained
from the inequality
\begin{equation}
|P^{\rm Yu}(a)|\leq \tilde{\Xi}(a),
\label{eq2}
\end{equation}
\noindent
where $[-\tilde{\Xi}(a),\tilde{\Xi}(a)]$ is the minimum confidence interval
\cite{12} containing all differences
$P^{\rm theor}(a)-\bar{P}^{\rm expt}(a)$ within the separation region
$180\,\mbox{nm}<a<746\,$nm. The Yukawa pressure between two parallel
plates is given by
\begin{eqnarray}
&&
P^{\rm Yu}(a)=-2\pi G\alpha \lambda^2\,{\rm e}^{-a/\lambda}
\nonumber \\
&&~
\times
\left[\rho_{\rm Au}-(\rho_{\rm Au}-\rho_{\rm Cr})\,
{\rm e}^{-\Delta_{\rm Au}^{\!(s)}/\lambda}\right.
\nonumber \\
&&~~~~~~~~~~~~~~~~~~
\left.
-(\rho_{\rm Cr}-\rho_s)\,
{\rm e}^{-(\Delta_{\rm Au}^{\!(s)}+\Delta_{\rm Cr}^{\!(s)})/\lambda}\right]
\nonumber \\
&&~
\times
\left[\rho_{\rm Au}-(\rho_{\rm Au}-\rho_{\rm Cr})\,
{\rm e}^{-\Delta_{\rm Au}^{\!(p)}/\lambda}\right.
\nonumber \\
&&~~~~~~~~~~~~~~~~~~
\left.
-(\rho_{\rm Cr}-\rho_{\rm Si})\,
{\rm e}^{-(\Delta_{\rm Au}^{\!(p)}+\Delta_{\rm Cr}^{\!(p)})/\lambda}\right].
\label{eq3}
\end{eqnarray}
\noindent
Here, $\Delta_{\rm Au}^{\!(s)}$, $\Delta_{\rm Cr}^{\!(s)}$,
$\Delta_{\rm Au}^{\!(p)}$, and $\Delta_{\rm Cr}^{\!(p)}$ are the
thicknesses of 
the Au and Cr layers with the densities $\rho_{\rm Au}$ and
$\rho_{\rm Cr}$ on the sphere and the plate, made of sapphire with density
$\rho_{s}$ and Si with density $\rho_{\rm Si}$, respectively (the values
of all parameters are contained in \cite{12,11}).

To verify the constraints obtained in \cite{12,11} from Eq.~(\ref{eq2})
based on the use of the simplified 
PFA for Yukawa forces, we return instead to the
original inequality following from the experimental data, i.e., that
the gradient of the Yukawa force normalized to $2\pi R$ belongs to the
confidence interval determined at a 95\% confidence level
\begin{equation}
\left|\frac{1}{2\pi R}\,
\frac{\partial F_{sp}^{\rm Yu}(a)}{\partial a}\right|\leq\tilde{\Xi}(a).
\label{eq4}
\end{equation}
\noindent
Here, $F_{sp}^{\rm Yu}(a)$ is the Yukawa-type force for the configuration
of a sapphire sphere and Si plate, both covered with Cr and Au layers.
Equation (\ref{eq4}) has never been used in previous literature
(including \cite{12,11,15b}) to obtain constraints on the
parameters of the Yukawa-type interaction from
the measurements of the Casimir force.
When dealing with the Yukawa interaction
for the experimental parameters of \cite{12,11},
the plate can be considered as
infinitely large. This was proved in \cite{14} and confirmed in \cite{1}.
The exact expression for  
$F_{sp}^{\rm Yu}$ acting between a homogeneous
sphere and a plane plate of sufficiently large area and thickness (as in
experiment \cite{12,11}) was found in Eq.~(6) of
\cite{15}.
Applying this equation to the layer structure of \cite{12,11},
one arrives at
\begin{eqnarray}
&&
F_{sp}^{\rm Yu}(a)=-4\pi^2 G\alpha \lambda^3\,{\rm e}^{-a/\lambda}\,
\left[
\vphantom{{\rm e}^{-\Delta_{\rm Au}^{\!(s)}/\lambda}}
\rho_{\rm Au}\,\Phi(R,\lambda)\right.
\nonumber \\
&&~~
-(\rho_{\rm Au}-\rho_{\rm Cr})\,
\Phi(R-\Delta_{\rm Au}^{\!(s)},\lambda)\,
{\rm e}^{-\Delta_{\rm Au}^{\!(s)}/\lambda}
\label{eq5} \\
&&~~
-(\rho_{\rm Cr}-\rho_s)\,
\Phi(R-\Delta_{\rm Au}^{\!(s)}-\Delta_{\rm Cr}^{\!(s)},\lambda)\,
\nonumber \\
&&~~~~~~~~~~~~~~~~~~~~~~~~~~~~~
\left.\times
{\rm e}^{-(\Delta_{\rm Au}^{\!(s)}+\Delta_{\rm Cr}^{\!(s)})/\lambda}\right]
\nonumber \\
&&~
\times
\left[\rho_{\rm Au}-(\rho_{\rm Au}-\rho_{\rm Cr})\,
{\rm e}^{-\Delta_{\rm Au}^{\!(p)}/\lambda}\right.
\nonumber \\
&&~~~~~~\left.
-(\rho_{\rm Cr}-\rho_{\rm Si})\,
{\rm e}^{-(\Delta_{\rm Au}^{\!(p)}+\Delta_{\rm Cr}^{\!(p)})/\lambda}\right.
\nonumber \\[1mm]
&&~~~~~~~~~~~~~~~\left.
-\rho_{\rm Si}\,
{\rm e}^{-(\Delta_{\rm Au}^{\!(p)}+\Delta_{\rm Cr}^{\!(p)}+
\Delta_{\rm Si}^{\!(p)})/\lambda}
\right],
\nonumber
\end{eqnarray}
where
\begin{equation}
\Phi(r,\lambda)\equiv\Phi^{\rm exact}(r,\lambda)=r-\lambda+
(r+\lambda)\,{\rm e}^{-2r/\lambda}.
\label{eq6}
\end{equation}
\noindent
Note that if we put
\begin{equation}
\Phi(r,\lambda)=\Phi^{\rm PFA}(r,\lambda)\equiv R,
\label{eq7}
\end{equation}
\noindent
Eq.~(\ref{eq5}) results in $F_{sp}^{\rm Yu,PFA}(a)$ which is also
expressible as
\begin{equation}
F_{sp}^{\rm Yu,PFA}(a)=2\pi R E^{\rm Yu}(a),
\label{eq8}
\end{equation}
\noindent
where $E^{\rm Yu}(a)$ is the Yukawa energy per unit area of two parallel
plates covered with thin layers as described above.
Note that Eq.~(\ref{eq8}) is nothing but the simplified
formulation of the PFA.
The negative derivative of $E^{\rm Yu}(a)$ with respect to $a$
results in the Yukawa pressure  between two 
parallel plates in Eq.~(\ref{eq3}).

Note that the thickness of the lower plate mentioned in the
explanations to Eq.~(\ref{eq8}) might be finite whereas the upper
plate is a semispace (see paper \cite{2}). This is because the
force $F_{sp}^{\rm Yu,PFA}(a)$ in the left-hand side of
Eq.~(\ref{eq8}) can depend only on the three geometrical
parameters: separation $a$, sphere radius $R$ and thickness of
the lower plate $D_1$ (if it is not much larger than $a$,
as in the experiments under discussion). However, Eq.~(6)
of \cite{1} which is analogous to our Eq.~(\ref{eq8}) (the
use of $P_{\rm Yu}$ instead of $E^{\rm Yu}$ is a misprint)
contains in the right-hand side the energy per unit area of
two parallel plates where the upper plate is of some
arbitrary finite thickness $D_2$. This equation is incorrect
because its left-hand side does not depend on $D_2$.
Using this incorrect formulation of the PFA, paper \cite{1}
performed an extensive investigation of the dependence
of $F^{\rm Yu,PFA}$ on $D_2$ [see Eq.~(9) and Figs.~2 and 3]
which is physically meaningless.

We next compare the constraints on the Yukawa parameters $\alpha,\,\lambda$
obtained in \cite{12,11} from Eqs.~(\ref{eq2}), (\ref{eq3}) using
the simplified PFA, 
and the exact results following from Eqs.~(\ref{eq4})--(\ref{eq6}).
According to \cite{1}, the larger $\lambda$ is, 
the greater is the error
in the constraints obtained using the PFA. The strongest constraints following
from the data of Refs.~\cite{12,11} were obtained within the interaction
range $20\,\mbox{nm}<\lambda<86\,$nm. Hence as the first example
we compare the resulting constraints at $\lambda=86\,$nm. For this fixed
$\lambda$ the strongest constraints follow at $a=250\,$nm where the
half-width of the confidence interval is $\tilde{\Xi}(a)=1.52\,$mPa \cite{12}.
From the exact Eqs.~(\ref{eq4})--(\ref{eq6}) we then obtain
\begin{equation}
\alpha^{\rm exact}=2.88167\times 10^{13}\approx 2.88\times 10^{13}.
\label{eq9}
\end{equation}
\noindent
Using the simplified 
PFA for the Yukawa interaction i.e., from Eqs.~(\ref{eq2}) and
(\ref{eq3}), it follows that
\begin{equation}
\alpha^{\rm PFA}=2.88011\times 10^{13}\approx 2.88\times 10^{13}.
\label{eq10}
\end{equation}
\noindent
Note that only the latter value was obtained in \cite{12}, and used
to plot the line separating the allowed and prohibited regions in the
($\alpha,\,\lambda$)-plane. It follows from a comparison of
Eqs.~(\ref{eq9}) and (\ref{eq10}) that the constraints on
$\alpha,\,\lambda$ obtained in \cite{12} are exact, to the
quoted level of precision.

As a second example, we consider the much larger value,
$\lambda=400\,$nm, in the
beginning of the region where the constraints obtained from the torsion
pendulum experiment are the most stringent \cite{14}. For this $\lambda$
the strongest constraints are obtained from the Casimir force data at
$a=400\,$nm and   the
half-width of the confidence interval is $\tilde{\Xi}(a)=0.45\,$mPa \cite{12}.
In this case the exact results in Eqs.~(\ref{eq4})--(\ref{eq6})
yield
\begin{equation}
\alpha^{\rm exact}=2.03189\times 10^{11}\approx 2.0\times 10^{11},
\label{eq11}
\end{equation}
\noindent
whereas Eqs.~(\ref{eq2}) and  (\ref{eq3}) using the 
simplified PFA for the Yukawa force
lead to
\begin{equation}
\alpha^{\rm PFA}=2.02708\times 10^{11}\approx 2.0\times 10^{11}.
\label{eq12}
\end{equation}
\noindent
The latter value was used in \cite{12} to plot
the line separating the allowed and prohibited regions in the
($\alpha,\,\lambda$)-plane.

Note that the half-width of the confidence interval $\tilde{\Xi}(a)$ is
found from the total theoretical and experimental errors in the Casimir
pressure determined at a 95\% confidence level. It takes into account
all constituent errors, including the random and systematic experimental
errors. Also included are
theoretical errors due to uncertainty of the optical data,
and due to the use of the PFA. As a quantity obtained from the errors,
$\tilde{\Xi}(a)$ is calculated to only two, or at maximum to three, significant
figures [the latter happens at shortest separations alone where
$\tilde{\Xi}(a)$ is relatively large]. It would thus be
an evidently inconsistent result  if the two
different determinations of the constraints
on ($\alpha,\,\lambda$) with two theoretical expressions for
$F_{sp}^{\rm Yu}(a)$ differing by only
0.1\% would lead to markedly different strengths of the
resulting constraints
(to say nothing of the constraints differing by 100\%, as discussed in
\cite{1}).

One can conclude that within the whole range of $\lambda$ considered in
\cite{12,11} the strength of the derived constraints is precisely the
same irrespective of whether or not the 
simplified PFA was used for the calculation
of the Yukawa-type interaction between the sphere and the plate.

The opposite conclusion mentioned above was arrived at in \cite{1}.
According to \cite{1}, exact computations of the
Casi\-mir force ``for the recent Casimir sphere-plane
experiment [21] (here \cite{11})... gives a deviation from
PFA of the order of 0.1\% at the smallest value of
$a/R\approx 0.001$ reached in the experiment
($a_{\rm min}\approx 160\,$nm). Since the limits
to non-Newtonian forces are obtained using the residuals in the Casimir
theory-experiment comparison, in order to meaningfully replace the exact
formula of the Yukawa force with its PFA approximation, the level of
accuracy between these two should be therefore a small fraction, for instance,
10\% of the accuracy with which the Casimir force is controlled by using
PFA rather than the exact expression for the sphere-plane Casimir force.
If this condition is not fulfilled, the derived limits could be
off also by a large order of 100\% correction. However, targeting a 10\%
accuracy level with respect to the Casimir theory-experiment accuracy
implies deviations from $\eta=1$ of 0.01\%, which can be obtained...
only in the range of $\lambda$ below 100\,nm. The presence of
substrates with different densities...'' leads to the situation
that $\eta$ ``in the case of the experiment reported in [21] (here
\cite{11}), is equal to 1.00126, i.e. a correction already
equal to 0.126\%.'' Thus, according to \cite{1}, the
deviation of the quantity $\eta$ from unity in the experiment
\cite{11} exceeds allowed deviations by a factor of 13 and this
may lead to order of 100\% correction to the obtained
constraints.

This statement is based on a simple misunderstanding. The key point is
that the relative error in the maximum allowed value of $\alpha$
obtained using the simplified 
PFA is determined by the sum of relative errors
in $\tilde\Xi(a)$ and in the application of the PFA to calculate the
Yukawa force. Keeping in mind that $\tilde\Xi(a)$ has a meaning of an
absolute error, it is determined with only two or three
significant figures independently of its value. From this it
follows that $\tilde\Xi(a)$ can only be known with the relative error
of about $0.5$\%. As can be seen from the above citations,
\cite{1} mistakenly links
a demand to the value of the relative
error introduced by the application of the PFA to
calculate the Yukawa force
(of about 0.01\%) with the magnitude of $\tilde\Xi(a)$ (or its
constituent part due to the application of the PFA to
calculate the Casimir force). Actually, to obtain constraints
on $\alpha$ with the relative error of about 1\% from
the data of any
high precision future experiment on the measurement of the Casimir
force, it is quite sufficient to calculate the Yukawa force
with a relative error of about 0.1\% \cite{16}. 
For example, at
$a=\lambda=400\,$nm this calculation can be safely performed
using the simplified formulation of
the PFA if the application conditions of this formulation
($a,\,\lambda\ll R,\,D_1$) are satisfied \cite{2}.
 Thus, the conclusion of \cite{1} that the limits derived
in \cite{12,11} could have ``a large order of 100\% correction'' is
invalid.

\section*{Acknowledgements}

E.F. was supported in part by DOE under Grant No.~DE-76ER071428.
G.L.K. and V.M.M. were  supported by
Deut\-sche Forschungsgemeinschaft, Grant No.~GE\,696/10--1.
\hfill\\
G.L.K.\ was also supported by the Grant of the Russian Ministry of
Education P--184.

\end{document}